# HOW PARTICLES CAN EMERGE IN A RELATIVISTIC VERSION OF BOHMIAN QUANTUM FIELD THEORY

T. Mark Harder

*E-mail address:* tmharder1@shockers.wichita.edu

It is shown how bosonic material particles can emerge from a covariant formulation of de Broglie-Bohm theory. The formulation is based on the work of Nikolic. Material particles are continuous fields, formed as the eigenvalue of the Schrodinger field operator, evaluated along a Bohmian trajectory.

The de Broglie/Bohm (dBB) Quantum Field Theory occupies a controversial position with respect to the interpretation of quantum phenomena. Our purpose in this note is not to argue in favor of this approach, but to exploit it as an arena for presenting a point of view regarding the wave/particle duality problem. The dBB theory makes some definite statements regarding: the vacuum state of quantum field theory, the classical limit and the measurement problem. We believe it has value for this reason, if only to introduce a different paradigm for examining certain plaguing fundamental problems, at the expense of introducing several new problems of its own. Recently it has been receiving attention in regard to other fundamental problems: the role of information in quantum theory, the proper venue for a theory of quantum gravity, and the perennial problem of how to interpret the quantum formalism itself. These introductory comments are all old news, and the formalism for this approach is very well explained in references [1,2]. We will follow this formalism, with the addition of a mass term in the Boson field equation. A modification of the formalism introduced by Nikolic in [3] will provide a crucial paradigm shift from the usual approach, and render the theory covariant. In a following paper, relating to the Fermion aspect of the problem, we will depart from Holland's presentation and utilize other techniques, also due to Nikolic. Some effort will be made to clarify the reasoning behind the two approaches as the argument progresses.

The approach followed here will utilize the following line of reasoning. The dBB theory is not Lorentz covariant. If one quantizes a classical field, the process of obtaining the eigenvalue (field) of a Schrodinger field operator does not yield a scalar function. Consider the quantum equation

$$\hat{\phi}(\mathbf{x})|\chi(\mathbf{x})\rangle = \chi(\mathbf{x})|\chi(\mathbf{x})\rangle. \tag{1}$$

This does not result in a scalar for the field $\chi$ in the sense that $\chi'(\mathbf{x},t) \neq \chi(\mathbf{x},t)$ if $x'^{\mu} = x'^{\mu}(x')$ is a Lorentz transformation. Also, the actual *particle* in the so-called dual solution does not play much of a role in the mathematical structure of the theory. In fact, there is no agreed-upon mathematical entity that is generally designated as *the particle* in the theory. One merely says the *particle* follows one of the *indicated* trajectories; the single object being that thing which actually penetrates one of the two slits (for example) while the guiding wave goes through *both* slits.

Our point of view here is that the particle is represented as the eigenvalue of the field $\chi$ evaluated along a field trajectory. Our formalism is quite commonplace, and relies upon a familiar construction. This is a manifestly complex representation for a scalar field, limited by a constraint. This approach makes the usual machinery more transparent and is almost always followed in the literature. The machinery of the dBB theory provides a non-linear field which will be the field we are seeking.

Let us begin with the introduction of MacKinnon's [4] solution to the Klein Gordon equation. A similar (but NOT identical) expression will arise in the construction of a model for a free scalar particle in the context of the Nikolic-dBB theory. Starting with the Klein-Gordon equation for mass $m$

$[\nabla^2 - c^{-2}\partial_t^2]\varphi = \mu^2 \varphi$, $\mu = mc/\hbar$    consider the expression

$$\varphi_m = \frac{\sin(\mu r)}{r} e^{-i\mu\gamma(ct-(v/c)z)}, \tag{2}$$

with $r = \sqrt{x^2 + y^2 + \gamma^2(z-vt)^2}$   $v =$ velocity   $\gamma = 1/\sqrt{1-\frac{v^2}{c^2}}$   and with $\mu$ a mass parameter. The MacKinnon representation above is taken from [5], and is a localized envelope moving at speed $v$ multiplied by a superluminal term. Both this and (2) are considered a single term moving in the z direction with velocity $v$. This function has been much studied with regard to de Broglie's "double solution"; in this respect, see [6] for a large collection of relevant papers. It is worth mentioning that, in spite of its strange appearance, the exponential in equation (2) reduces to $e^{i(\omega t - kz)}$ when the kinematical relations $m\gamma v =$ momentum;   $\omega = c\sqrt{k^2 + m^2}$   are used. In addition, the relations

$\frac{\partial S}{\partial t} = -$ Energy;   $\frac{\partial S}{\partial z} =$ momentum  $(\hbar = 1)$ with $S$ the phase, appear to be true (but are not). Equation (2) is a particular formalism for describing a localized structure, and thus applies to any entity that obeys the Klein Gordon equation. Here we are not particularly interested in the actual model for the particle, as in a consistent description how such a model would be included in the conventional formalism.

Equation (2) may be easily derived in the following way. If one writes $\varphi_m = f(\rho,z)e^{-i\mu ct}$ as a solution to the Klein Gordon equation, the function $f$ obeys $[\nabla_\perp^2 + \frac{\partial^2}{\partial z^2} + \mu^2]f(\rho,z) = 0$. The function f then immediately obeys $f(\rho,z) = \frac{\sin(\mu\sqrt{\rho^2+z^2})}{\sqrt{\rho^2+z^2}}$. Performing a Lorentz boost in $z$, on $\varphi_m$, results in equation (2).

## De Broglie Bohm Quantum Field Theory

De Broglie-Bohm QFT starts in the Schrodinger picture with an operator $\hat{H}$ acting on a wave function $\Psi(\phi(\mathbf{x}),t) = \langle \phi(\mathbf{x}) | \Psi(t) \rangle$, which is a functional of the real field coordinate $\phi$ and a function of the time $t$. The formulation is well covered in references [1,2,3] and we will follow that formulation. Bohm theory begins with writing $\Psi([\phi(\mathbf{x})],t) = R([\phi(\mathbf{x})],t)e^{iS([\phi(\mathbf{x})],t)/\hbar}$ with $R([\phi(\mathbf{x})],t)$ and $S([\phi(\mathbf{x})],t)$ real functionals. In Holland's and dBB theory, the guidance formula

$$\frac{\partial \phi(\mathbf{x},t)}{\partial t} = \frac{\delta S[\phi(\mathbf{x}),t]}{\delta \phi(\mathbf{x})}\bigg|_{\phi(\mathbf{x})=\phi(\mathbf{x},t)} \tag{3}$$

yields the time evolution of the field $\phi$. The equation satisfied by the field coordinate $\phi$ is found by applying $\delta/\delta\phi$ to the equation satisfied by $S$. If we identify $\delta S/\delta\phi$ with $\dot\phi$ and define

$$\frac{d}{dt} = \frac{\partial}{\partial t} + \int d^3x \frac{\partial \phi}{\partial t}\frac{\delta}{\delta\phi}, \text{ then we get } \partial^\mu\partial_\mu\phi + \frac{\partial V(\phi)}{\partial\phi} = -\frac{\delta Q[\phi(\mathbf{x}),t]}{\delta\phi(\mathbf{x})}\bigg|_{\phi(\mathbf{x})=\phi(\mathbf{x},t)}, \tag{4}$$

a nonlinear equation.

## De Donder-Weyl Formalism of Bohmian Mechanics

References [1,2] develop several basic scenarios for the dBB theory. Some puzzling results are: the vacuum is not a Lorentz covariant concept; equation (3) for the field does not yield a Lorentz scalar, (which often leads to the speculation that it may be the actual particle "dual solution" sought by de Broglie); field values at any instant are non-locally connected to other values. The important feature of Nicolic's work is the achievement of a relativistic framework for the dBB theory; and the formulation of a relativistic guidance formula

$$\partial^\mu \phi = \frac{dS^\mu}{d\phi}. \tag{5}$$

[NB this follows from the classical equation of motion (12) will then be written $\partial^\mu \phi = \frac{\partial S^\mu}{\partial \phi}$ ]  (6)

Here $d/d\phi$ generalizes the ordinary partial derivative $\partial/\partial\phi$ to nonlocal functionals. An example of a functional that is nonlocal in space but local in time is $\frac{\delta A([\phi], x')}{\delta \phi(x)} = \frac{\delta A([\phi], x')}{\delta \phi(\mathbf{x}; x^0)} \delta(x'^0 - x^0)$.  (7)

Nikolic postulates the equation $\quad \frac{1}{2} \frac{dS_\mu}{d\phi} \frac{dS^\mu}{d\phi} + V + Q + \partial_\mu S^\mu = 0.$  (8)

Note in this equation that $S^\mu = S^\mu([\phi], x)$ is a functional of $\phi(x)$ and thus that $S^\mu$ at $x$ may depend on the field $\phi(x')$ at all points $x'$. In [3] the argument is also taken the other way, and conventional Bohm theory is derived from the covariant version. This demands some additional relations. These involve $S$. $S^0$ must be local in time; and $S^i$ must be completely local, so that $dS^i/d\phi = \partial S^i/\partial\phi$. This leads to $dS^i = \partial S^i + \partial_t \phi \, dS^i/d\phi$. Another result is given by $\partial^\mu \phi = dS^\mu/d\phi$, which is a covariant version of (3).

**EMERGENCE OF PARTICLES**

We will now revert to the notation utilized in [1, 2]. That is, we will use an expansion of the field coordinates into the normal modes of a box with volume $V = L^3$,

$$\varphi(\mathbf{x}) = V^{-1/2} \sum_{\mathbf{k}} q_{\mathbf{k}} e^{i\mathbf{k}\cdot\mathbf{x}}.$$  (9)

The $q_{\mathbf{k}}$ are complex numbers with the reality condition $q_{-\mathbf{k}} = q_{\mathbf{k}}^*$, to preserve reality of $\phi$, and $k^i = \frac{2\pi n^i}{L}$, $i = 1, 2, 3$ with $n^i$ being a positive or negative integer, including zero. The inverse transformation to (9) is $\quad q_{\mathbf{k}} = V^{-1/2} \int_V \varphi(\mathbf{x}) e^{-i\mathbf{k}\cdot\mathbf{x}}.$  (10)

The functional derivatives become $\quad \frac{\delta}{\delta\varphi(\mathbf{x})} = V^{-1/2} \sum_{\mathbf{k}} e^{-i\mathbf{k}\cdot\mathbf{x}} \frac{\partial}{\partial q_{\mathbf{k}}}$ and $\frac{\partial}{\partial q_{\mathbf{k}}} = V^{-1/2} \int_V e^{-i\mathbf{k}\cdot\mathbf{x}} \frac{\delta}{\delta\varphi(\mathbf{x})}.$  (11)

In these coordinates Schrodinger's equation becomes $\quad i \frac{\partial \Psi}{\partial t} = \sum_{\mathbf{k}/2} \left( -\frac{\partial^2}{\partial q_{\mathbf{k}} \partial q_{\mathbf{k}}^*} + k^2 q_{\mathbf{k}} q_{\mathbf{k}}^* \right) \Psi$  (12) where the notation **k**/2 recognizes that **k** and –**k** modes are identical. Finally, the creation and annihilation operators are represented $a_{\mathbf{k}}^{\dagger} = \frac{1}{\sqrt{2\omega_{\mathbf{k}}}}\left(kq_{\mathbf{k}}^{*} - \frac{\partial}{\partial q_{\mathbf{k}}}\right)$ and $a_{\mathbf{k}} = \frac{1}{\sqrt{2\omega_{\mathbf{k}}}}\left(kq_{\mathbf{k}} + \frac{\partial}{\partial q_{\mathbf{k}}^{*}}\right)$, and the equation of motion is

$$\frac{\partial q_{\mathbf{k}}}{\partial t} = \frac{\partial S}{\partial q_{\mathbf{k}}^{*}}\bigg|_{q_{\mathbf{k}}=q_{\mathbf{k}}(t)}. \tag{13}$$

From the remarks following (8), this (Holland's) notation is consistent if $S^0$ is local in time. The equation for the vacuum is $\Psi_0[q,q^*,t] = N\prod_{\mathbf{k}/2}\exp(-\omega_{\mathbf{k}}q_{\mathbf{k}}^{*}q_{\mathbf{k}} - iE_0 t)$, $\omega_{\mathbf{k}}^2 = \mathbf{k}^2 + \mu^2$. (14)

In (14), $E_0$ is the total energy, quantum plus "classical" potential energy.

Let us now take advantage of the covariant formalism and solve equation (13), in the proper frame, where the "particle" is at rest. Applying the creation operator to the vacuum for a single mode we get $\Psi_{\mathbf{k}} = \sqrt{2\omega_{\mathbf{k}}} q_{\mathbf{k}}^{*} e^{-\omega_{\mathbf{k}} t}\Psi_0$. More generally, a range of modes may be taken to represent a one quantum state and so we will write

$$\Psi_1[q,q^*,t] = \sum_{\mathbf{k}} f_{\mathbf{k}} \sqrt{2\omega_{\mathbf{k}}} q_{\mathbf{k}}^{*} e^{-\omega_{\mathbf{k}} t}\Psi_0 \tag{15}$$

taking the $f$'s to be (possibly) complex constants.

Starting with equation (15), let us pursue the following line of reasoning. If the particle is at rest, we will take $\omega_{\mathbf{k}} = \mu$, and let all the $f$'s be the same constant; i.e. a *standing* wave. Define the set $M = \{\mathbf{k} \mid \mathbf{k}\cdot\mathbf{k} = \mu^2\}$. For the standing wave, the only k's that contribute will be those in $M$. For the phase we will get $S = \frac{1}{2i}\log\left(\sum_{\mathbf{k}\in M} f_{\mathbf{k}} q_{\mathbf{k}}^{*} \bigg/ \sum_{\mathbf{k}\in M} f_{\mathbf{k}} q_{\mathbf{k}}\right) - \omega_{\mathbf{k}} t - \sum_{\mathbf{k}/2}\omega_{\mathbf{k}} t$. (16)

Hence, using (13), and since all the q's are the same, $\dot{q}_{\mathbf{k}} = \frac{\partial S}{\partial q_{\mathbf{k}}^{*}} = \frac{1}{2iq_{\mathbf{k}}^{*}}$ for each k. (17)

Before proceeding, let us briefly address the issue of a *relativistic* trajectory. In the scalar case one defines the trajectory from $\frac{dx^{\mu}}{d\tau} = \frac{j^{\mu}}{2m\phi^*\phi}$, for a one particle wave-function $\phi$. The affine parameter can be eliminated by $\frac{d\mathbf{x}}{dt} = \frac{\mathbf{j}(t,\mathbf{x})}{j_0(t,\mathbf{x})}$ with $t = x^0, \mathbf{x} = (x_1, x_2, x_3), \mathbf{j} = (j_1, j_2, j_3)$. In field theory, this equation is written in the Schrodinger picture in terms of the wave function $\varphi(\mathbf{x},t) = \int D\varphi \Psi_0^*[\varphi]\varphi(\mathbf{x})\Psi[\varphi,t]$. In this language the trajectory equation will become $\frac{d\mathbf{x}}{dt} = \frac{\varphi^*(\mathbf{x})\vec{\nabla}\varphi(\mathbf{x})}{\varphi^*(\mathbf{x})\vec{\partial}\varphi(\mathbf{x})}$. These may all be easily generalized for the many particle case. Now if the field representing the particle is a plane wave, and the kinematical relations $m\gamma v =$ momentum; $\omega = c\sqrt{k^2 + m^2}$ are used, then for rectilinear motion at constant speed, one may use the usual Lorentz transformation.

Let us solve equation (17). If one assumes the solution is $q_\mathbf{k} = Ae^{-i\omega_\mathbf{k} t}$ then $A = \frac{1}{\sqrt{2\omega_\mathbf{k}}}$. Let $k \equiv \sqrt{\mathbf{k} \cdot \mathbf{k}}$.

We can write $\varphi$ as $\varphi = \frac{1}{\sqrt{V}}\sum_\mathbf{k} e^{i\mathbf{k}\cdot\mathbf{x}} \delta_{k,\mu} \int_0^t q_\mathbf{k} dt' = \frac{1}{\sqrt{V}}\sum_\mathbf{k} e^{i\mathbf{k}\cdot\mathbf{x}} \delta_{k,\mu} \int_0^t \frac{1}{2}(q_\mathbf{k} + q_\mathbf{k}^*) dt'$

$$= \frac{1}{\sqrt{V}}\sum_\mathbf{k} e^{i\mathbf{k}\cdot\mathbf{x}} \delta_{k,\mu} \frac{1}{\sqrt{2\omega_\mathbf{k}}} \int_0^t \cos(\omega_\mathbf{k} t') dt';$$

$$= \frac{1}{\sqrt{V}}\sum_\mathbf{k} e^{i\mathbf{k}\cdot\mathbf{x}} \delta_{k,\mu} \frac{1}{\sqrt{2\omega}} \theta(t)\frac{\sin(\omega t)}{\omega}$$

where we let $\omega_\mathbf{k} = \mu$. The frequency factor in the exponential $q_\mathbf{k} = Ae^{-i\omega_\mathbf{k} t}$ is taken to be the same as the classical value, unlike the equivalent problem presented in references [1,2]. This maintains consistency and also yields a relativistically invariant result.. Now noting that

$$\left(\frac{1}{2\pi}\right)^2 \int \frac{\delta(\mu-k)}{k} e^{i\mathbf{k}\cdot\mathbf{r}} k^2 \sin\theta \, dk\, d\theta\, d\varphi = \left(\frac{1}{2\pi}\right)\int \delta(\mu-k) \frac{e^{ikr\cos(\theta)}}{ir\sin(\theta)}\Big|_0^\pi \sin\theta \, dk$$

$$= \frac{-1}{\pi}\int \delta(\mu-k) \frac{e^{-ikr} - e^{ikr}}{2ir} dk = \frac{1}{\pi}\int \delta(\mu-k) \frac{\sin(kr)}{r} dk$$

$$= \frac{1}{\pi}\frac{\sin(\mu r)}{r}$$

and making the association $\frac{1}{V}\sum_\mathbf{k} = \frac{1}{(2\pi)^3}\int d^3\mathbf{k}$ the final result for the field is

$$\varphi = \frac{1}{\sqrt{2V\omega_\mu}} \frac{1}{2\pi^2} \frac{\sin(\mu r)}{r} \theta(t)\sin(\omega_\mu t) \quad \text{where } r = \sqrt{x^2 + y^2 + z^2}. \tag{18}$$

Since this was carried out in the "particle's" rest frame, and our solution is now an actual scalar, we may boost this in the z direction and get the result

$$\varphi = \frac{1}{\sqrt{2V\omega_\mathbf{k}}} \frac{1}{2\pi^2} \frac{\sin(\mu\sqrt{\rho^2 + \gamma(z-vt)^2})}{\sqrt{\rho^2 + \gamma(z-vt)^2}} \theta(\gamma(ct-v/cz))\sin(\omega_\mathbf{k}\gamma(ct-v/cz)) \tag{19}$$

where we have inserted the speed of light c for clarity and $\gamma = 1/\sqrt{1 - \frac{v^2}{c^2}}$, $\omega_\mathbf{k} = \sqrt{k^2 + m^2}$.

**DISCUSSION**

Note that in the rest frame, both the spatial and temporal factors behave as delta functions for large $t, r$. In fact, (19) resembles a Helmholtz solution (Bessel function) multiplied by a temporal green function. In light of reference [7] the solution is a continuous field, save at $t = 0$, of infinite extent, and the particle is a spherical standing wave when it is at rest.

The Fermion analog of this one body problem will have some special difficulties, but should proceed similarly. Reference [8] outlines the procedures for Fermion fields in the context of the dBB theory. One would observe that each component of the spinor solution to Dirac's equation solves the Klein Gordon equation and use procedures similar to the above. Of course, the formulation would not be that simple; to render the calculation similar one must use the formulation of reference [9] wherein it is possible to view the wave functionals as overlaps with Grassmann field states. Also reference [10] points out how to Bosonize a Fermionic theory to describe it in terms of observables.

We have taken a simplistic view of the trajectory equations for quantum field theory, which is the reason that we have only discussed a constant velocity one body problem. It is anticipated that a rigorous discussion of entangled states within the context of a formulation such as this would prove interesting.

There are other covariant versions of dBB theory, [see 11]. The one we have picked is a bit more avant-garde in that it requires a special foliation of space-time, but the foliation emerges dynamically as a quantum effect; and we find this remarkably satisfying. Also, it would appear to offer promise for a venue to achieve a theory of quantum gravity.

A relativistic formulation of dBB does not come without a price. The vacuum is still a non-covariant concept; in the sense that there exists a special reference frame that can be formulated as in equation (14). Thus one affirms there exists a frame wherein everything is at rest. From this frame, one may induce Lorentz transformations such that the dynamics of the theory are covariant. Then the claim is made that there do not exist conditions such that one can experimentally determine the static frame. This is somewhat like the formalism that allows Newtonian gravitation to be formulated in a generally covariant manner, where one introduces quantities defined to be identical in all reference frames.

## ACKNOWLEDGEMENT

The present author had a fruitful interchange with the author of [3, 8, and 10] on a preliminary version of this paper. Professor Nikolic was kind enough to point out some egregious errors in our original ideas. The current formulation is our own, and we are alone responsible for the errors contained therein.